%% file: infocom.tex
\documentclass[10pt, conference, letterpaper]{IEEEtran}
\pdfoutput=1
\usepackage[space,nocompress]{cite}
\usepackage{amsmath,amssymb,amsfonts}
\usepackage{graphicx}
\usepackage{textcomp}
\usepackage{xcolor}
\usepackage{siunitx}
\usepackage{xspace}
\usepackage{multirow}
\usepackage{xcolor,microtype}
\usepackage{xspace,relsize}
\usepackage[hyphens,lowtilde]{url}
\usepackage{dsfont}
\usepackage{booktabs}

\DeclareMathOperator*{\argmaxA}{arg\,max}
\renewcommand{\vec}[1]{\boldsymbol{#1}}
\usepackage{mathtools}
\usepackage{algorithm}
\usepackage{algorithmicx}
\usepackage[noend]{algpseudocode}

\newcommand*\Let[2]{\State #1 $\gets$ #2}

\newcommand{\para }[1]{
	\subsubsection{#1}
}
\newcommand{\1}{{\em (i)}}
\newcommand{\2}{{\em (ii)}}
\newcommand{\3}{{\em (iii)}}

\newcommand{\reminder}[1]{{\color{blue}\textsf{[#1]}}}
\newcommand{\note}[1]{{\color{red}{\bf #1}}}

\renewcommand{\reminder}[1]{}\renewcommand{\note}[1]{}

\newcommand{\zipphone}{ZipPhone\xspace}

\newcommand{\mpl}{{\sf PhoneLab}}
\newcommand{\phl}{$\mpl$\xspace}
\newcommand{\mrm}{{\sf RealityMining}}
\newcommand{\rem}{$\mrm$\xspace}

\begin{document}

\title{\zipphone: Protecting user location privacy\\ from cellular service providers}
\author{Keen Sung\\\texttt{ksung@cs.umass.edu} \and Brian Levine\\\texttt{brian@cs.umass.edu} \and Mariya Zheleva\\\texttt{mzheleva@albany.edu}}
\maketitle

\input{abstract}
\input{introduction}
\input{related}
\input{overview}

\input{attacker_model}
\input{dynamics}
\input{classifiers}
\input{evaluation}

\input{zipphone}
\input{discussion}
\input{conclusion}
\bibliographystyle{IEEETran}
\bibliography{mobi}

\end{document}

%% file: abstract.tex
\begin{abstract}
Wireless service providers track the time and location of all user connections. 
Location inference attacks have been effective in revealing the identity of anonymous users of wireless services. 
In this paper, we propose \zipphone, a solution that leverages existing cellular
infrastructure to improve user privacy.
Spartacus allows a community of users to strategically time their connections to remain anonymous while incurring a minimal loss of utility.
We evaluate \zipphone from the perspective of a cell service provider and a community of privacy-seeking users, and quantify the privacy/utility trade-off of \zipphone using two datasets containing cell tower logs of hundreds of users. We present and assess a deanonymization algorithm that uses both \textit{location profiling} and \textit{trajectory linking}.
We find that by renewing identifiers every ten minutes and remaining offline for 30 seconds, users can reduce their identifiability by up to 45\%.
\end{abstract}


%% file: introduction.tex
\section{Introduction}

There are more than 5 billion subscribers to mobile services worldwide~\cite{Stryjak:2019}. 
When these users connect to the Internet, service providers such as 
Verizon and Google Fi 
typically store a log of the time, radio tower, and user identity associated with the connection, to manage the network and account for billing~\cite{Zang:2011}. Accordingly, the approximate location of each user over time is known as well. 
It has always been necessary for mobile users to trust their provider with location information because it is necessary to authenticate to towers for network service. 4G and 5G technologies make use of base stations that are more densely deployed, each covering a more limited range than 2G and 3G networks. Therefore, as cellular networks have advanced, providers have  learned with finer resolution where authenticated users are located. 
The increased location specificity not only allows for more accurate pinpointing of a user's location, but also makes it easier to identify the user of a device using \emph{location profiling}~\cite{Mulder:2008} methods; these methods exploit users' historical pattern of movement to identify them.

In this paper, we address a fundamental problem between users and service providers:
the user seeks connectivity for a mobile device, and privacy for their location history, but the provider requires that users authenticate and pay for service before a device is connected. Existing methods for increasing user privacy are generally focused on the network or application layers, and they do not help users that desire location privacy from a cellular service provider.  For example,  VPNs and Tor~\cite{Dingledine:2004} mask the  IP address of a user from a remote web site, and they also thwart the cellular operator from learning the IP address of the remote site; but they do not protect a user from the provider learning their location or history of movements. Users may also wish privacy from location based services (LBS), but again, obfuscation of queries to a remote server does not address privacy from the mobile operator~\cite{niu2015enhancing}. A recent class action lawsuit~\cite{eff-att} demonstrates that mobile users desire that cellular service providers not sell their historic movement records to third parties, such as {\em location aggregators}.


We propose a method called \emph{\zipphone} that allows celluar operators to provide  service to authenticated, paid users without requiring the users  to reveal  their identity, thereby protecting their location and history of geographical movements as well. Our evaluations show that our solution can thwart location profiling attacks with only a small loss  in user utility, and that power costs are minimal. \zipphone provides anonymous authenticated connectivity based on ephemeral mobile identifiers that thwart linking of movement by one device. In sum, \zipphone provides privacy by separating the purchase of credentials from their use on the network. Further, 
\zipphone can be deployed by mobile network operator (MNO) (e.g.,  Verizon) or by
a mobile virtual network operator (MVNO) (e.g., Google Fi). Either way, the user does not need to trust the MNO or MVNO. \zipphone makes use of existing 3GPP standards for roaming and {\em eSIMs}; the hurdles required for deployment are related to the provider's business model rather than software or hardware upgrades. 


\zipphone provides users with pseudonymous credentials to connect to the network. However, this is not enough: users can reveal their identity by visiting a unique location usch as their home. Once the credential is compromised, past and future movements by the user are as well.
Moreover, past work has demonstrated that some users exhibit diurnal and otherwise predictable movement behaviour~\cite{Zang:2011}, and that an attacker can use location profiling from historical records to deanonymize users. Users can leverage the well known concept of mix zones~\cite{Beresford:2003} to minimize the consequences of having one location discovered. With \zipphone, users switch to a new identifier regularly enough to naturally form ephemeral mix zones which can preserve anonymity. 

We primarily evaluate an attacker's ability to identify a user by using both {\em location profiling} and {\em trajectory linking} in combination. 
We further show that an attacker's ability to identify a user depends on the {\em predictability} and amount of {\em mixing}
in each user's behaviour; e.g., users who exhibit regular movement patterns that are distinct from other users are most easily identifiable. We also quantify the loss of utility for \zipphone users defending against such attacks. 

%
Using datasets collected from real users~\cite{Eagle:2006,nandugudi2013phonelab}, we determine the privacy gained from small deployments of \zipphone with 100--150 users in the same locale using uncoordinated mix zones. Despite hiding within only a small number of users, we show that privacy gains are significant, especially for users who are not unique but are highly predictable; a larger deployment would increase privacy while maintaining each user's utility and performance. 

Our contributions are organized as follows.

\begin{itemize}
	
	\item We outline the \zipphone protocol and the fundamental parameters that affect its privacy in Section~\ref{sec:attacker_model}.
	
	\item We introduce a sophisticated location privacy attack to reveal the identity of anonymous cell tower traces. We 
	probabilistically link traces together to more accurately identify users using knowledge of users' historical movement patterns. In Section~\ref{sec:attack}, we discuss this attack and a strategy to avoid it.
	
	\item We evaluate our approach on two data sets collected from real cellular users~\cite{Eagle:2006,nandugudi2013phonelab} in Section~\ref{sec:eval}. 
Predictable, mixing users are identifiable 24\% of the time if they renew their identifiers every ten minutes, as opposed to 69\% if they use a long-term pseudonym. We quantify the trade off between the frequency of renewals and user utility. We find that hourly renewals offer little protection, but that renewals of about 10 minutes offer significant gains. 
	\item Also in Section~\ref{sec:eval}, we estimate the incurred battery use from  \zipphone for 3G and 4G networks. Specifically, we measured power consumption during network association and disassociation, and we estimate that a user may incur at most 1\% battery overhead per day regardless of network technology or desired privacy if \zipphone were used.
	
	\item In Section~\ref{sec:lte}, we detail the LTE and 5G-compatible signaling and communications for \zipphone that allows anonymous operation via an MNO or MVNO that accepts cryptocurrency for payment of services. 
\end{itemize}


%% file: related.tex
\section{Related Work}\label{sec:related}

In contrast to related work, our goal is to provide mobile users location privacy from the cellular service provider itself. Existing work on mobile location privacy  has different goals, including: \1 properly anonymizing mobility datasets before public release; \2 adding privacy for users of locations based services; and \3 increasing location privacy for mobile device users from third-party attackers but {\em not} the service provider itself. A number of privacy studies that do assume an untrustworthy service provider
either do not handle location privacy, or require substantial changes to infrastructure.
We discuss the most relevant of these studies below.

Works that aims to prevent leaks in personally identifiable information in 
shared or publicly released datasets~\cite{yin2015re} primarily rely on obfuscation. These works also include efforts to prevent trajectory recovery~\cite{tu2018new,gramaglia2017preserving}.
Methods like differential privacy~\cite{dwork2011differential,mir2013dp} are used to add noise to data while preserving its aggregate characteristics. Older work on deanonymization of private traces of mobile users assume the user's pseudonym is unchanged throughout the trace. But a small amount of external information, such as the person's home or work address~\cite{Isaacman:2011}, can deanonymize an obfuscated trace~\cite{Beresford:2003,Beresford:2004,Krumm:2007,Mulder:2008,Golle:2009,Ma:2010} given a consistent identifier. Zang and Bolot~\cite{Zang:2011} show that suitably anonymizing a trace of 25 million cellular users across 50 states (30 billion records total) requires only that users have the
same pseudonym for no longer than a day. A day's duration is  unsuitable for Zang and Bolot's goal of supporting researchers that wish to characterize the behaviour of users over time (while maintaining their privacy). On the other hand, the result is promising for users seeking privacy, who might be able to change their pseudonyms much more frequently than once per day.

Work that increases the privacy of location-based services (LBS)~\cite{huang2018location,wang2018protecting,wang2018privacy,niu2015enhancing} generally add noise to location queries. 
These works are not viable or applicable against an untrusted service provider:  a user cannot  manipulate which tower they connect to, and the provider knows the physical locations of the towers serving  users.

Several studies protect against  third party attackers and vulnerabilities in 3GPP implementations~\cite{hong2018guti,hussain2019insecure}.  Khan et al.~\cite{khan2018defeating} provide a cryptographic mechanism to generate LTE pseudonyms and prevent third-party attackers or IMSI catchers from linking users.

We do not assume that an MNO will deploy any changes required to add privacy; however, many past works do assume MNOs will accommodate privacy mechanisms. Some focus on enlisting a trusted carrier to protect against a third party~\cite{Federrath:1996,Gorlatova:2011,Gorlatova:2011a,Foo:2012}. Reed et al.~\cite{Reed:1998} propose privacy from the carrier using onion routing, but does not consider the direct connection that must be made to a tower. Federrath et al.~\cite{Federrath:1995} propose a similar scheme that prevents linkability of calls between two parties but omit critical details regarding authentication to the carrier. Fatemi et al.~\cite{Fatemi:2010} propose an anonymous scheme for UMTS using identity-based encryption, but unlike our approach, their scheme involves the carrier in the cryptographic exchange; they enumerate the vulnerabilities of similar works~\cite{Park:2001,Jiang:2006,Yang:2005,Zhu:2004}. Kesdogan et al.~\cite{Kesdogan:1996} proposes using a trusted third party to create pseudonyms for GSM users, but also routes all calls through that provider, which allows it to characterize the calling pattern and infer the caller. 

Mix zones can be employed by a user against a provider attacker, unlike more recent location privacy techniques which rely on their cooperation. Our work is closely related to work on mix zones~\cite{Beresford:2004,Freudiger:2009} and abstention 
from service~\cite{Bindschaedler:2012}. 
Other work involves the introduction of false information\cite{Shokri:2011,Kido:2005}. 
Few studies use this concept to protect the user from an omnipresent network attacker. Chan~\cite{chan2015anoncall} focuses on call metadata privacy, rather than location privacy. Emara and Wolfgang~\cite{emara2015caps} use a mixing scheme to prevent physical trajectory linking. 

In comparison to related work, we differ in that we do not trust the MNO (or MVNO) to ensure the user's privacy, and we assume in our analysis that the adversary is attempting to link together traces. Our evaluations are based on traces of real users~\cite{Eagle:2006,nandugudi2013phonelab}, which allows us to quantify the periodicity of identifier changes in the context of modern cellular infrastructure.

%% file: overview.tex
\section{Overview and Attacker Model}
\label{sec:attacker_model}

Our primary design goal is to provide location privacy for cellular users from the carrier itself  using only mechanisms that are compatible with deployed MNO and MVNO  systems. 
In this section, we provide a high-level overview of \zipphone, which applies the classic concept of a mix zone~\cite{Beresford:2004} to LTE networks. We then define our attacker model.

\subsection{\zipphone Overview}
\label{zipphone}





\zipphone users all employ the same protocol, but they do not coordinate or communicate with one another; it is not a peer-to-peer system. We refer to {\em mix zones}, but users do not trade or mix IMSIs or other credentials --- all IMSIs and credentials are nonces used once by exactly one user. Furthermore, our zones are not set up as static geographic areas. 

For clarity and generality, we assume \zipphone is offered by an MVNO, however, nothing about our design prevents the MNO itself from offering \zipphone.

\zipphone users begin by anonymously paying for and obtaining eSIM {\em profile} credentials from an MVNO before joining the MNO's network, a process we detail in Section~\ref{sec:lte}. The profiles allow the user's device to join the MNO network using a short-lived IMSI, MSISDN, and other identifiers. \zipphone has the device follow a attach-and-detach cycle. The attach period always lasts for a minimum duration and detaching is only an option immediately after a new tower is attached to. The detach duration is selected at random from a uniform distribution. The device then reattaches with a new profile (and IMSI etc.), and the  cycle repeats. This algorithm is defined in Section~\ref{sec:attack}. The user is not involved in the protocol; the decision of when to detach/re-attach is handled by the device. The user does not need to alter her movements. If the device is actively being used, the \zipphone protocol will postpone detachment; and if the user were to actively use the device while detached, it will re-attach immediately.   While attached, the device uses  data services to support voice/phone  communication, and all data services must be via an anonymous communication system. Using the MNO's network, they can purchase additional eSIMs from the MVNO if necessary.  

The mechanisms for deploying \zipphone are relatively straightforward --- the challenge for \zipphone is in understanding its performance guarantees as a function of user behaviour dynamics. How does the duration of the on-off periods affect user privacy in the context of attacker strategies? 
We tackle this question first, in Sections~\ref{sec:attack} and~\ref{sec:eval}.
Then, in Section~\ref{sec:lte}, we present the technical details behind \zipphone including required signaling for integration with LTE networks, and the method used to purchase service anonymously from a provider. 


%% file: attacker_model.tex


\subsection{Problem Statement}
In current LTE systems, a user  $u$ is assigned a permanent IMSI identifier $i$  and joins the network.  The user attaches to a sequence of towers as it moves according to signal strength, creating a {\em trace} associated with the IMSI: $(i, (s_1, s_2,\ldots ))$, where each value of $s$ indicates a specific tower and a timestamp.  The MNO knows the mapping of users to IMSIs: $u: i$, etc.

\zipphone users seek to use the network, but not have their real identities associated with their traces.  The goal of the attacker is to infer and label their identities from the traces. 
It is not the goal of the user to hide that they are using \zipphone.


We assume the attacker is an MNO that already has a history of traces for each \zipphone user. The attacker's goal is to determine which user from a set $u_1, u_2, u_3, \ldots$ is the one that  created the trace  $(i, (s_1, s_2,\ldots ))$ based on a classifier trained from the known history. 

In Section~\ref{sec:results}, we demonstrate that longer traces are easier to identify and link with other traces; users should regularly renew their identifier in order to keep these traces short. 
We assume the user does not  perturb their own  movement patterns. Therefore important parameters are  \1 the renewal frequency, and \2 the offline duration. When the renewal frequency is higher, privacy also increases; but each renewal incurs an offline period and increases power usage. Longer offline durations reduce linking but reduce utility. We assume all such parameters are public and known to the attacker.


%
 
\zipphone makes use of ephemeral IMSI identifiers managed in software, instead of permanent IMSI values normally assigned to SIM cards. This approach is more practical and effective than several naive solutions. To some extent, users could repeatedly purchase ``burner phones", but as our results show, 
even changing the identifier once every few hours is not sufficient to thwart an attacker with a historic profile of a user. 

\subsection{Attacker Model}

We assume the attacker  \1 has all traces of all \zipphone devices, and \2 has labelled/identified  traces of historic movement for all \zipphone users, for training a classifier; in other words, the attacker is an MNO. The attacker  performs \emph{trajectory linking}, which patches together separate traces if a classifier predicts they are from the same user.  

We assume that all \zipphone users are of equal interest to the attacker, and that it uses only normal cellular infrastructure to attack. For example, we assume that the MNO attacker does not install cameras on towers to identify users via facial recognition, nor would they follow a particular user by car. It does not make sense for the attacker to set up an IMSI catcher\cite{Dabrowski:2014} since they already own the entire real infrastructure. 




We assume that the MNO gains no other information from the users; this is feasible, since advertisement of other identifiers (e.g. IMEI, device model, or OS signatures) is easily turned off via OS settings. 
In practice, such features would assist the attacker (see~\cite{Corner:2017}), but are not the focus of this paper as they are more easily obfuscated or falsified than real geographical movement. For example, IMEIs, which are akin to a MAC address,  can be modified by the user since she controls the handset hardware (e.g., SilentCircle's blackphone~\cite{silentcircle}). Users are likely identifiable by the unique set of outgoing calls they make; however, they can make calls via VOIP through an anonymizing proxy or circuit rather than using the cellular carrier.  Encryption of the VOIP stream can thwart carrier eavesdropping. Stronger protection is available by using VOIP over Tor~\cite{torfone}.

%% file: dynamics.tex
\section{Attacker-defender dynamics}
\label{sec:attack}

In this section, we define the exact algorithms used by the ZipPhone user and the MNO-based attacker. 

\MakeRobust{\Call}
\begin{algorithm}[t]
\caption{User identifier renewal strategy}\label{alg:user}
\begin{algorithmic}[1]
\Let{utility}{Minimum utility between 0.0 and 1.0}
\Let{max\_off\_time}{Maximum time offline during renewal}

\While {device is online}
    \State \Call{wait}{until device moves outside range of tower}
    \State \Call{disconnect}{}
    \State off\_time $\gets$ \Call{uniform}{0,max\_off\_time}
    \State \Call{wait}{off\_time}
    \State \Call{connect}{}\Comment{connect with new identifier}
    \State cooldown\_time $\gets$ utility $\times$ off\_time
    \State \Call{wait}{cooldown\_time}
\EndWhile
\end{algorithmic}
\end{algorithm}

\subsection{User strategy}

Algorithm~\ref{alg:user} defines the ZipPhone user algorithm. As described in the previous section, ZipPhone users renew their identifiers when: \1 they are in the process of switching towers, and \2 the renewal  \emph{cool down period} (in seconds) has expired; \3 they are not actively using the phone. To renew, users first detach,  then stay offline, and then reattach with a new profile. The offline time is selected uniformly at random from a \emph{maximum offline period} (our evaluations set the maximum to 30s). It must be random, otherwise linking traces would be trivial.  The cool down period ensures that utility remains at a minimum for the user. This aggressive renewal strategy is frequent enough to allow the natural formation of mix zones, and does not require users to coordinate times or places to mix.  








%% file: classifiers.tex

\subsection{Attacker strategies}
\label{sec:classifiers}


\begin{algorithm}[t]
\caption{Location profiling algorithm}\label{alg:profiling} 
\begin{algorithmic}[1]
\Function{profile\_user}{$u$}\Comment{$u$ is the user index}
    
    \State $T^u_{0,q} \gets \frac{Count(q)}{\sum_{q'\in \mathds{C}} Count(q')}$ \Comment{The prior for user's initial location}
    \ForAll{$p\rightarrow q \in \Call{transitions}{u}$} \Comment{$p\rightarrow q$ denotes a transition}
        \State $T^u_{p,q} \gets \frac{Count(p \rightarrow q )}{\sum_{{q'}\in \mathds{C}}{Count(p
        \rightarrow {q'})}}$ \Comment{This transition matrix may be sparse}
    \EndFor
    \State \Return{$T^u$}
\EndFunction
\Function{classify\_user}{$\vec{s}$}\Comment{$\vec{s}=(s_0,s_1\dots), s\in \mathds{C}$ is a sequence of tower IDs}
    \State \Return $\arg\max_u T^u_{0,s_0} \prod_{i=0}^{n-2} T^u_{s_{i},s_{i+1}}$
\EndFunction
\end{algorithmic}
\end{algorithm}

The attacker's goal is to take a timestamped sequence of visited towers and infer the user, given a training set. We first describe a \emph{location profiling classifier} (Algorithm~\ref{alg:profiling}) that could be employed by the attacker. We then define a \emph{trajectory linking classifier} (Algorithm~\ref{alg:linking}) to aid the attacker in location profiling.

\subsubsection{Location profiling algorithm}
\label{profiling_classifier}

Our classifier is a Markov model that chooses the most likely user for a sequence of tower attaches; the classifier is adapted from Mulder et al.~\cite{Mulder:2008}. 
Vector $\vec{s}$ is a sequence of IDs in tower set $\mathds{C}$: $\vec{s}=(s_0,s_1,s_2\dots), s\in\mathds{C}$.
In the steps below, the attacker identifies the most probable user given each candidate user's history, $\hat{u}=\arg\max_u p(u|\vec{s})$.

\begin{align*}
\shortintertext{We determine the most likely user, given a sequence of locations.}
\Pr(u|\vec{s}) &= \Pr(u|s_0,s_1,s_2,\dots)\\
\shortintertext{We apply Bayes' rule, and consider the likelihood of a sequence given a user.}
\Pr(u|\vec{s}) &= \frac{\Pr(s_0,s_1,s_2,\dots|u)\Pr(u)}{\Pr(s_0,s_1,s_2,\dots)}\\
\shortintertext{We assume that each user is equally likely.}
\Pr(u|\vec{s}) &\propto \Pr(s_0,s_1,s_2,\dots|u)\\
&= \Pr(s_0|u)\cdot\Pr(s_1|u,s_0)\cdot\Pr(s_2|u,s_0,s_1)\cdot\\&\quad\Pr(s_3|u,s_0,s_1,s_2)\dots\\
\shortintertext{Each transition is independent per the Markov assumption.}
&= \Pr(s_0|u) \prod_{i=0}^n \Pr(s_{i+1}|s_{i},u)\\
\shortintertext{We determine the most likely user $\hat{u}$.}
\hat{u} &= \argmaxA_u \Pr(s_0|u) \prod_{i=0}^n \Pr(s_{i+1}|s_{i},u)
\end{align*}

The attacker computes a transition matrix $T$ for each user in the training data by counting the occurrences of 
these transitions in history. The probability of the first location in the sequence $\Pr(s_0|u)$ is computed
from the overall number of a user's occurrence at a location. The attacker does not consider the probability of a trace
ending at a certain location, since a sequence can end for arbitrary reasons.

The success of such an attack depends on two factors: the number of users in the same vicinity as the anonymous target, and the similarity of the user's location trajectory to those of the surrounding users. If there is one registered cell phone user on the network, then linking the user to location is trivial; however, if there are many users who behave similarly, it would be difficult for the attacker to tell the user apart. 

(We also defined and tested a classifier that added diurnal features but it did not perform significantly better.)

\algdef{SE}[DOWHILE]{Do}{doWhile}{\algorithmicdo}[1]{\algorithmicwhile\ #1}%
\begin{algorithm}[t]
\caption{Linking algorithm}\label{alg:linking} 

\newcommand{\lu}{{\textrm{l}}}
\newcommand{\rightarrowmax}{\xrightarrow{\text{max\_t}}}
\begin{algorithmic}[1]
\Let{max\_t}{Maximum time offline during renewal}
\Function{train\_link\_transitions}{}
    \ForAll{$p\rightarrowmax q$}\Comment{all locations $q$ seen within max\_t of $p$}
        \State $T^\lu_{p,q} \gets \frac{Count(p \rightarrowmax q )}{\sum_{{q'}\in \mathds{C}}{Count(p
        \rightarrowmax {q'})}}$ \Comment{transition matrix used for linking}
    \EndFor
    \State \Return $T^\lu$
\EndFunction

\Function{classify\_user\_with\_trajectory}{$\vec{s}$}
    \While{link\_count$<$max\_links}
        \State candidates $\gets$\Call{find\_candidates}{$\vec{s}$}\Comment{traces $\leq$ max\_off\_time after $\vec{s}$ ends}
        \If{\Call{empty}{candidates}}
            \State break
        \EndIf
        \State $\hat{\vec{s}'} \gets \arg\max_{\vec{s}'} T^\lu_{\vec{s}_{n},\vec{s}'_{0}}$
        \Comment{$\forall s' \in$ candidates}
        \State $\vec{s}\gets$ \Call{concatenate}{$\vec{s}$,$\vec{s}'$}
    \EndWhile
    \State \Return \Call{classify\_user}{$\vec{s}$}
\EndFunction
\end{algorithmic}
\end{algorithm}

\subsubsection{Trajectory linking algorithm}
We extend the above algorithm to account for the attacker's ability to do this linking. The attacker uses the publicly available offline time to inform a semi-Markov \textit{linking transition matrix}. This is similar to the user transition matrices above, except it is time-limited to the offline time, so that unreasonable transitions do not confuse the classifier, and any unseen transitions occurring within that time frame are accounted for.

Our trajectory linking algorithm first searches for candidate traces that \textit{start} within the 
maximum offline time. If a sufficient number of traces start within the offline time, 
the targets have a chance to mix, thus reducing the attacker's likelihood of classifying and increasing their 
privacy. However, the attacker may still infer the links between several traces using the linking transition matrix, and attempt to classify the resulting longer trace. Algorithm~\ref{alg:linking} extends the location profiling 
to include these attempts to link.




%% file: evaluation.tex

\section{Evaluation}\label{sec:eval}

In this section, we evaluate Algorithms~\ref{alg:user}, \ref{alg:profiling}, and~\ref{alg:linking} using two real-world datasets that contain geotagged user data coupled with tower attachment logs: \phl~\cite{nandugudi2013phonelab} and \rem~\cite{Eagle:2006}. Our datasets are, thus, similar to what would be available to an MNO attacker. In particular, we evaluated the amount of privacy a user could attain given certain sacrifices in utility. 

Before evaluating our algorithms, we characterize the amount of \emph{predictability} and \emph{mixing} behaviour exhibited by users in these datasets. We demonstrate that both characteristics are related to the success of the attacker's accuracy.

\subsubsection{Datasets} Both datasets were collected primarily from university affiliates who carried phones instrumented with software to log phone network attachment and activity. 

\begin{enumerate}
    \item {\phl}~\cite{nandugudi2013phonelab} is an Android testbed comprising 593 phones distributed to students at the University of Buffalo campus. As a part of this testbed, users contributed geotagged traces of their cellular network associations. We use 24 months from January 2015 to January 2017 of cellular network association traces from \phl to assess the privacy preservation potential of \zipphone.

\item \textbf{\rem}~\cite{Eagle:2006} is a dataset released by MIT that tracks a group of 100 mobile phone users across various contexts. Similar to \phl, \rem contains geotagged network association information. For our analysis, we leverage 12 months of \rem data from July 2004 to July 2005.

\end{enumerate}
We are unaware of other datasets that could be used to analyze our algorithms. (We filed IRB protocol 2017-3900 as part of this project, and it was approved as exempt.)

\input{data}
\subsection{Results}\label{sec:results} 
\input{results}

%% file: data.tex

\newcommand{\specialcell}[2][c]{%
	\begin{tabular}[#1]{@{}c@{}}#2\end{tabular}}
\begin{figure}[t]
\hspace{-10pt}
\relsize{-1}
\begin{tabular}{cccccc}
\toprule
\multirow{2}{*}{\textbf{Type}} & \multicolumn{2}{c}{\textbf{Trait}}    & \multirow{2}{*}{\specialcell{\textbf{Privacy}\\ \textbf{hypothesis}}} & \multirow{2}{*}{\textbf{PhoneLab}} & \multirow{2}{*}{\specialcell{\textbf{Reality}\\ \textbf{Mining}}} \\ 
                               & \textbf{Predictable} & \textbf{Mixing} &                                              &                                    &                                          \\\midrule
 P/M                         & Yes                  & Yes             & Moderate-Low                                     & 18\%                                & 18\%                                       \\ 
 nP/M                         & Yes                  & No              & Low                                          & 26\%                                & 30\%                                       \\ 
 P/nM                         & No                   & Yes             & High                                         & 30\%                                & 24\%                                       \\ 
 nP/nM                         & No                   & No              & Moderate                                 & 26\%                                & 29\%                                       \\ 
\bottomrule
\end{tabular}\smallskip
\caption{User typology and their proportions in our target datasets, with a hypothesis about the amount of privacy a user could attain from \zipphone.}\label{tab:usertypes}
\end{figure}

\subsection{Behaviour that affects attacker accuracy}
We begin by characterizing user behaviour. 
Intuitively, there are two behavioural traits that affect mobile users' privacy: \1~\textit{Predictability}, or to what extent users travel over fixed routes; and \2~\textit{Mixing} behaviour, or how likely are users to visit popular locations that see a large volume of other \zipphone users. To highlight the effect of user behaviour on privacy, we categorized \phl and \rem users {\em post hoc} into four groups: predictable or unpredictable; and mixing or not. The four resulting user types are described in Figure~\ref{tab:usertypes}, where we also set forth a hypothesis of how user behaviour would affect privacy. We verify and confirm these hypothesis in our evaluation (Section~\ref{sec:eval}).



\para{Predictability} We calculate the user predictability in terms of the similarity of the set of cellphone towers they visited during the testing and training period. For each user, let $C_{pre}$ be the set of towers visited during the training phase and $C_{post}$ be the set of towers visited in the testing phase. We express the predictability in terms of a user's Jaccard similarity score  $C_{pre}$ and $C_{post}$, defined as $J_C=\frac{C_{pre} \cap C_{post}}{C_{pre} \cup C_{post}}$, where $0 \leq J_C \leq 1$. $J_C=0$ when the sets of visited towers in testing and training are completely disjoint, while $J_C=1$ means that the sets of visited towers in testing and training are the same. Intuitively, a higher $J_C$ means a more predictable trajectory. 

\begin{figure}[t]
    \hfill
    \includegraphics[width=0.23\textwidth]{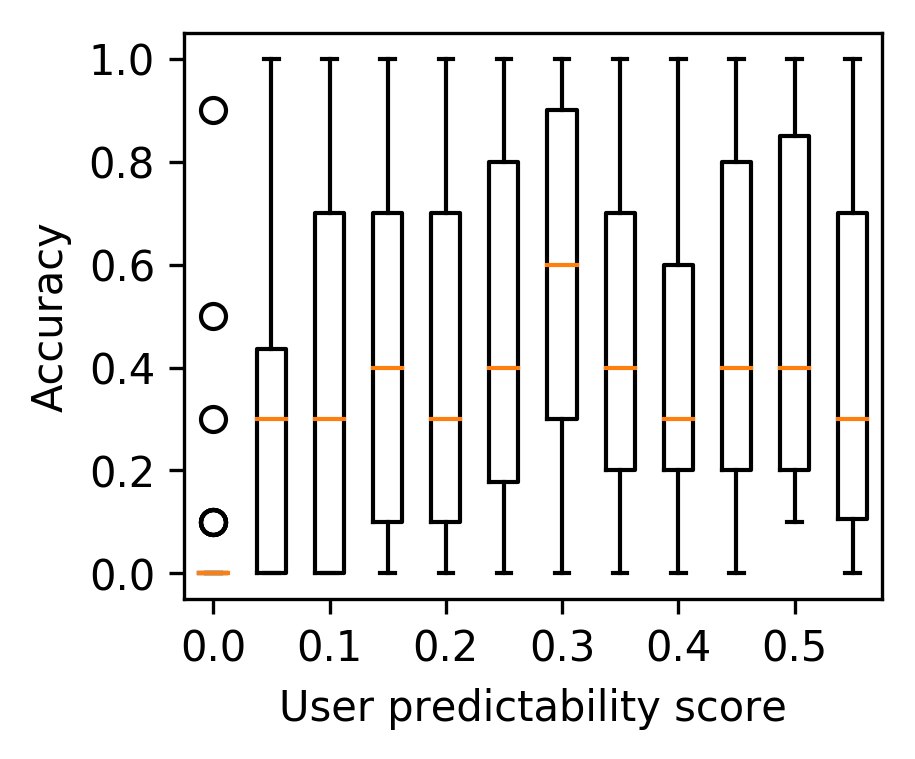}\hfill
    \includegraphics[width=0.23\textwidth]{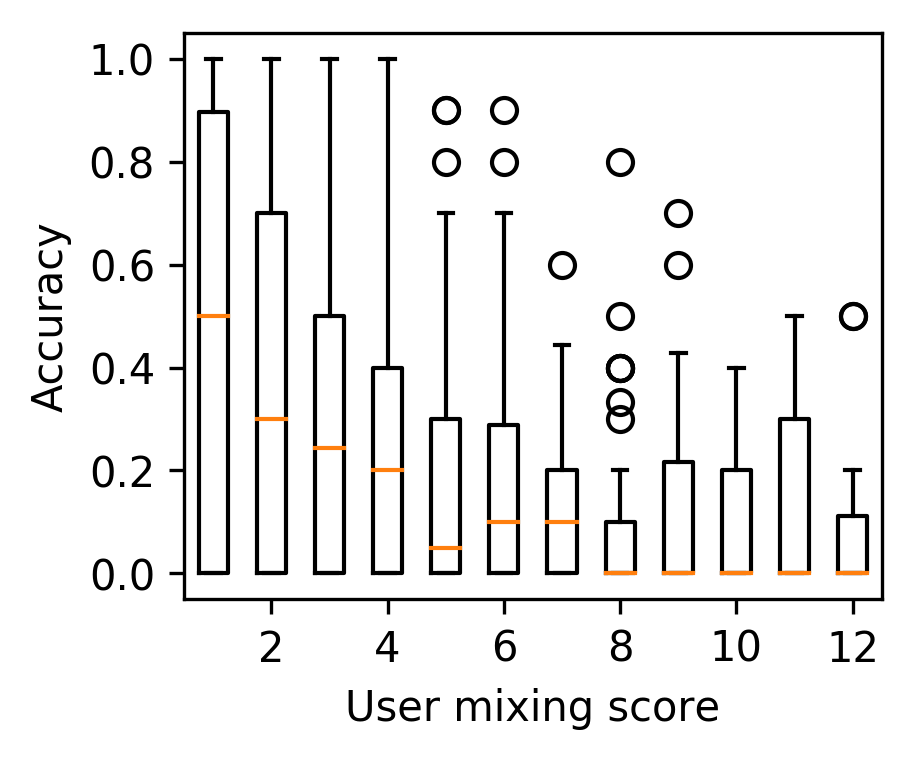}\hfill
    \caption{\textbf{Left:} User predictability versus attacker  accuracy, showing that attacker accuracy is near zero with low predictability. \textbf{Right:} User mixing versus attacker accuracy, showing that median accuracy falls to zero as user mixing increases. The plots were computed from the \phl dataset. The presented results are for a maximum offline time period of 30 seconds and a set utility of 95\%. Utility and accuracy metrics are discussed in detail in Section~\ref{sec:results}.}\label{fig:user_typology}
\end{figure}

Figure~\ref{fig:user_typology} (left) presents the attacker's accuracy (i.e., the probability that a user would be identified) as a function of the users' Jaccard score in the \phl dataset. We note that the trends and respective thresholds are similar for the \rem dataset and omit these results due to space limitations. For this setup, 91\% of users fall within the 0.0--0.4 Jaccard score range. For these users, we see an increasing trend in the attacker's accuracy as the Jaccard score grows. Using this analysis of our test dataset, we set the Jaccard score to 0.1 as a cut off to differentiate between predictable users (such with $J_C>0.1$) and unpredictable users (such with $J_C \leq 0.1)$.  

\para{Mixing behaviour} We establish a mixing score $\mathcal{M_C}$ as a metric that evaluates a user's likelihood to mix with other \zipphone users. Intuitively, the higher the mixing score, the more efficient ID switching will be and the harder it will be for the adversary to evade a user's privacy. We calculate $\mathcal{M_C}$ for each individual user. Let $t_i^k$ be the time a user $i \in (1,N)$ spends at tower $k \in (1,K)$. During the period $t_i^k$, other users $j \in (1,N'), j \neq i, N' \subset N$,  may arrive and depart from tower $k$. Let $\tau_{ij}^k$ be the time of user $j$'s arrival or departure. Intuitively, $t_i^k$ and $\tau_{ij}^k$ define the temporal granularity of tower mobility and \zipphone user encounter events, respectively, from the perspective of a single user $i$. Let $C(\tau_{ij}^k)$ be the number of users in user $i$'s vicinity at time $\tau_{ij}^k$. We define the mixing score as:
\begin{equation}
    \mathcal{M_C} = \sum_{k=1}^K \sum_{j=1}^{N'} \frac{C(\tau_{ij}^k)}{\tau_{ij}^k - \tau_{i(j-1)}^k}
\end{equation}
%
%
%
\noindent Figure~\ref{fig:user_typology} (right) presents the attacker's accuracy as a function of the users' mixing score in the \phl dataset. The trends and respective thresholds are similar for the \rem dataset. We see that the attacker's accuracy deteriorates as the users' mixing score increases. Based on this analysis, we set a mixing score of 4 as the cutoff to determine whether a user is mixing or not mixing, i.e. users with $\mathcal{M_C} \leq 4$ are not mixing and these with $\mathcal{M_C}>4$ are mixing.

\noindent\textbf{User typology in our datasets.} As detailed earlier, we differentiate between four types of users based on their predictability and mixing behaviour. Using the presented analysis in Figure~\ref{fig:user_typology}, we set a Jaccard similarity threshold of $0.1$ and mixing score threshold of $4$. Figure~\ref{tab:usertypes} presents the amount of users that fall in each user type category. We see a relatively even user representation across all categories. We use these user types and the corresponding user populations in all results presented in the evaluation of \zipphone (Section~\ref{sec:eval}). 

%% file: results.tex
To determine the affect of \zipphone on the utility and privacy of users, we simulated 
the protocol  using the \phl and \rem datasets.
In these simulations, the attacker uses the inference algorithms outlined in Section~\ref{sec:classifiers} to develop a location profile for each user based on two months of training data, and labels anonymous traces in the next month. 

\subsubsection{Utility-privacy trade-off}
\begin{figure}[t]
    \hfill
    \includegraphics[width=0.23\textwidth,trim={.25cm 0 0 0},clip]{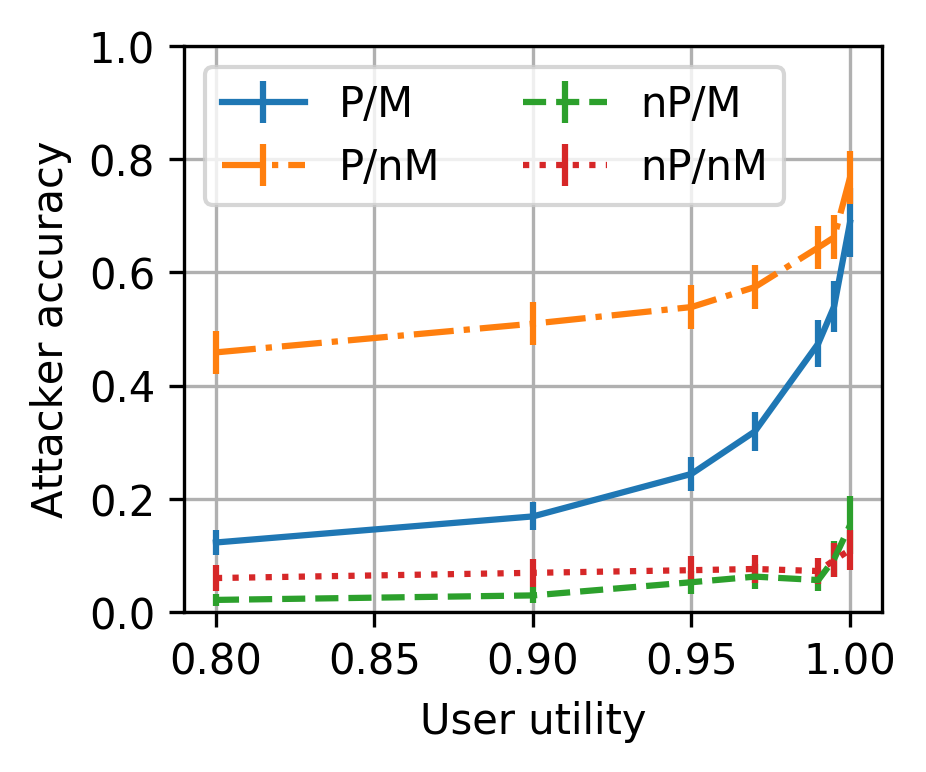}\hfill
    \includegraphics[width=0.23\textwidth,trim={.25cm 0 0 0},clip]{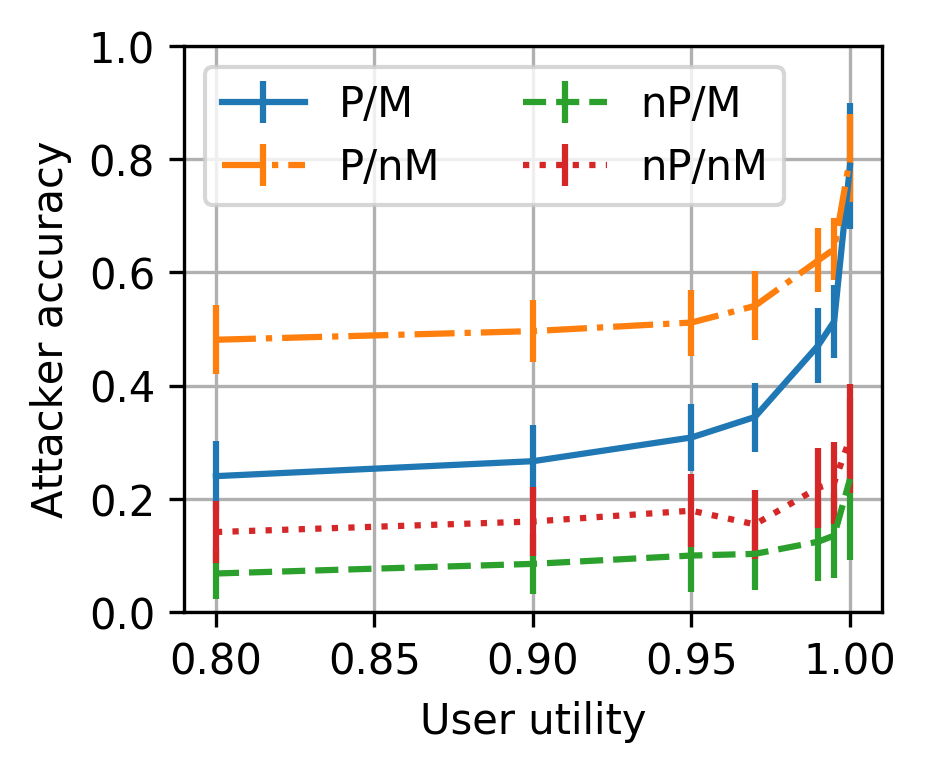}\hfill
\vspace{-1em}
    \caption{\textbf{Left}: \phl. \textbf{Right}: \rem.
    In both datasets, predictable but mixing users (Type P/M) gain the most from using \zipphone. Ten test traces were evaluated per user, and accuracy is represented as a mean of the proportion of successful reidentifications per user. Error bars represent a 95\% confidence interval.}  \label{fig:linked_results}\vspace{-1em}
\end{figure}

We evaluated the utility-privacy tension with regard to the four user types. We quantify privacy gained in terms of reduced attacker accuracy. We measured loss of utility  in terms of time spent offline during the testing period.
Figure~\ref{fig:linked_results} displays the privacy gained by each user group during the one-month testing periods.


Users gained significant privacy from sacrificing 5\% utility, on average remaining online for 9.5 minutes, and going offline for 30 seconds. In particular, Type P/M (predictable but mixing users) gained 45\% in the \phl dataset, and 49\% in the \rem dataset. Interestingly, Types nP/M and P/nM also show a similar trend: Type nP/M benefits from having the divided traces be less predictable, and for Type P/nM  any small amount of predictability is reduced to none. Type nP/nM does not mix, and enjoys uniformly high privacy because they are unpredictable. Users were more private in general in the \phl, since it represented a larger community of users, making mixing easier for the user, and user inference more difficult for the attacker.

\subsubsection{Trace length and location profiling}

\begin{figure}[t]
    \hfill
    \includegraphics[width=0.23\textwidth,trim={.25cm 0 0 0},clip]{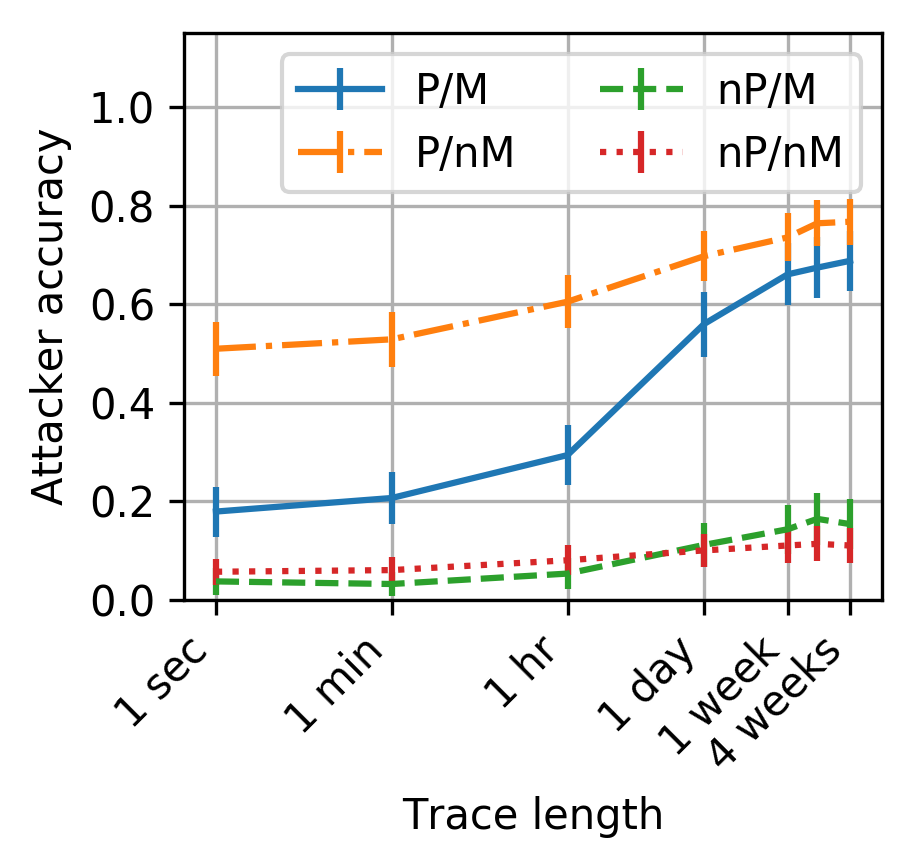}
    \hfill
    \includegraphics[width=0.23\textwidth,trim={.25cm 0 0 0},clip]{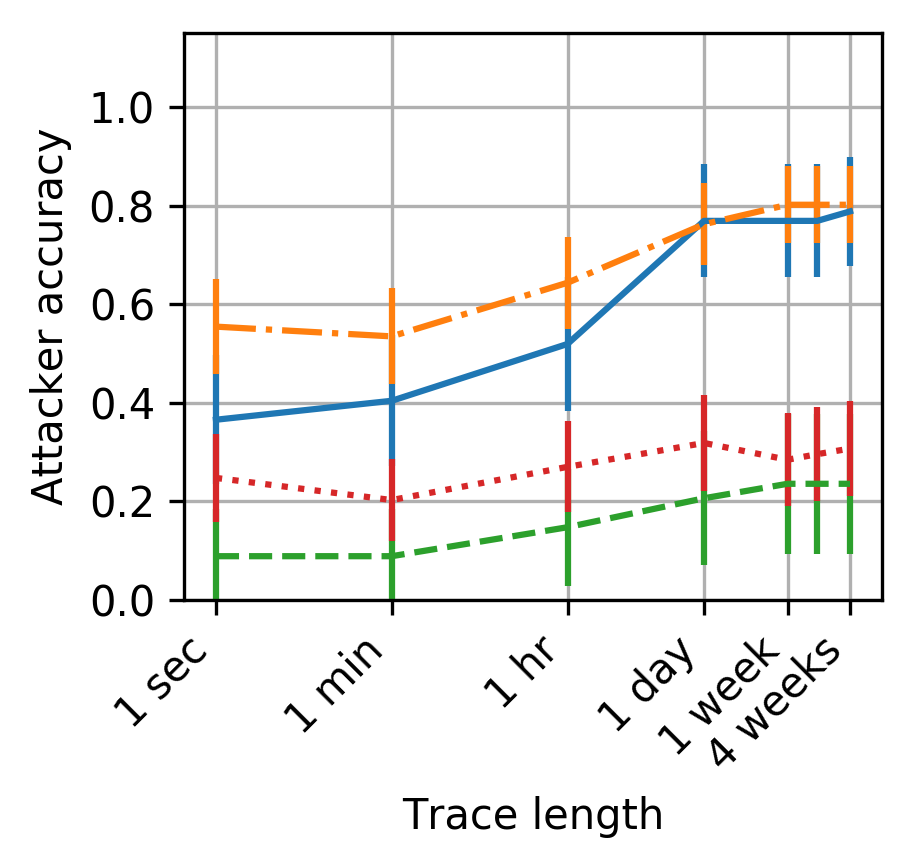}
    \hfill \vspace{-1em}
    \caption{\textbf{Left}: \phl. \textbf{Right}: \rem. Users lose a significant amount of privacy when traces are on the order of one day long. The accuracy at one month is equivalent to the accuracy in Figure~\ref{fig:linked_results} at 100\% utility.}\vspace{-1em}
    \label{fig:privacy_length}
\end{figure}

The main driver of attacker accuracy is trace length. In these experiments, the attacker tries to identify an independent trace of varying length. Figure~\ref{fig:privacy_length} shows the result.

In general, the longer the trace, the more identifiable (and thus less private) an individual is. Users who exhibit more predictable behaviour have less privacy; generally, they benefit from traces that are at most one hour long. In other words, predictable users should change their identifier at least once per hour while in motion. 
Those who travel to unique locations as compared to others benefit significantly less from the shorter trace.

This result highlights the benefit of \zipphone. Users should change their identifiers more than once per hour, and this system obviates the need to physically change an identifier, and handles this process automatically. While a temporary SIM device may grant some measure of privacy, a system that renews a user's identifier a lot more quickly can be a lot more effective.


\subsubsection{Compromises in utility}

\begin{figure}[t]
    \hfill
    \includegraphics[width=0.23\textwidth]{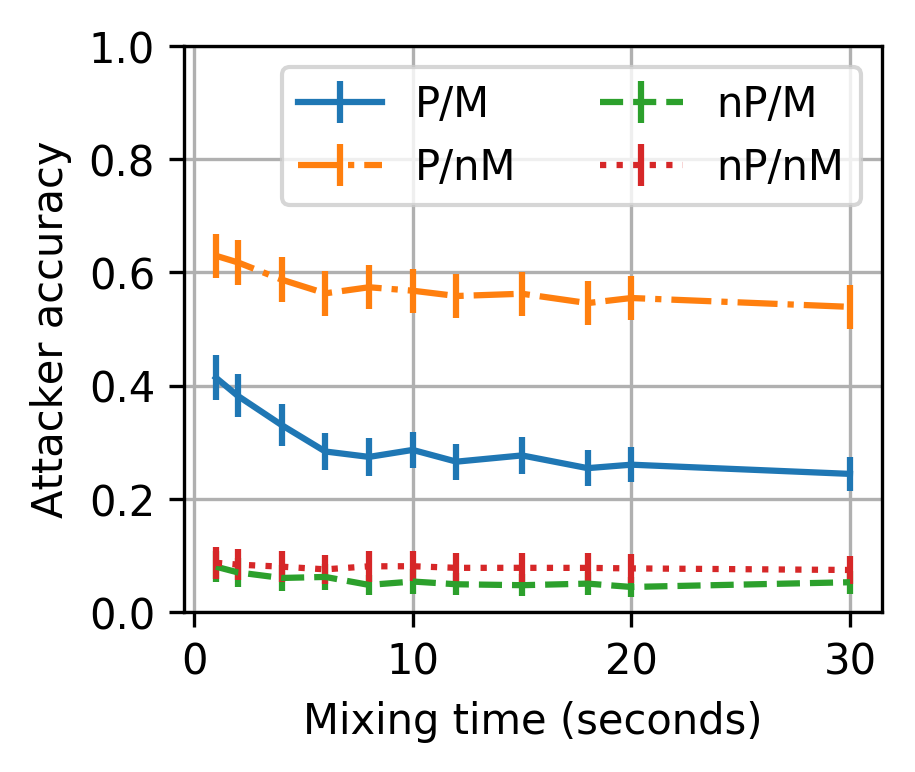}
    \hfill
    \includegraphics[width=0.24\textwidth]{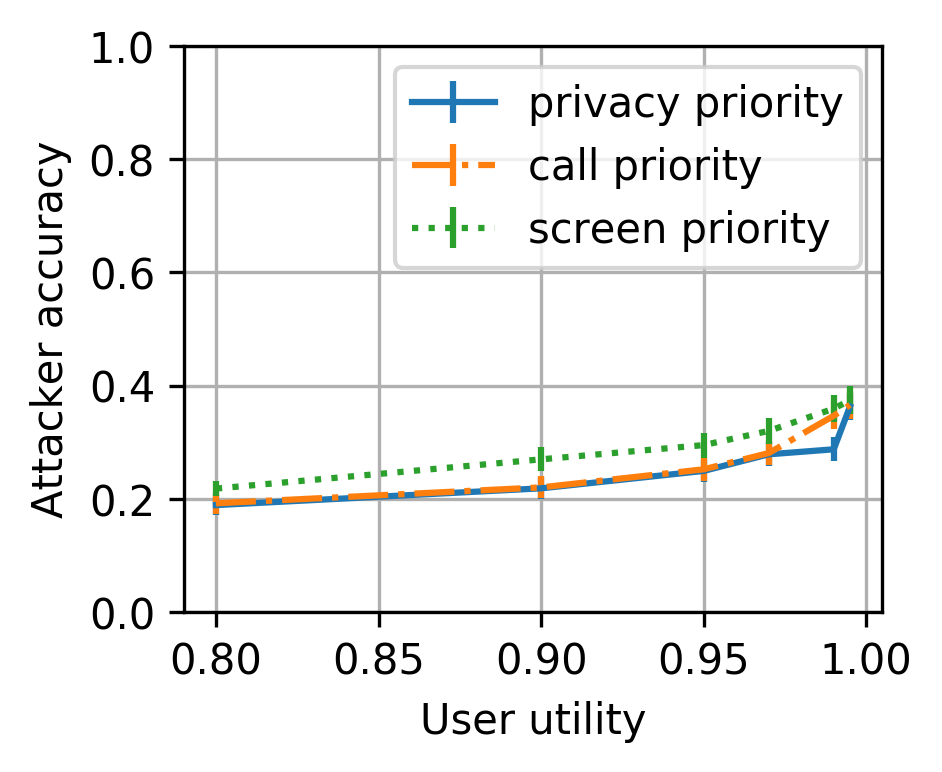}
    \hfill
    \caption{\textbf{Left}: the effect of mixing-time on privacy while maintaining a 95\% utility for the \phl dataset.
    \textbf{Right}: privacy/utility of all users depending on whether their priority is privacy, phone calls, or screen use. Calls can be prioritized without sacrificing privacy. However, remaining online while the screen is on significantly reduces privacy. 
    }
    \label{fig:dark_times}
\end{figure}

While users may renew identifiers by prearranging mixing strategies with other users, such coordination is impractical. A frequent enough renewal strategy and long enough renewal times allow mix-zones naturally form which allow users to mix without any coordination. In Figure~\ref{fig:dark_times}~(left), we examine the amount of time a user should remain offline. The frequency of renewal is informed by the utility desired, which we set at 95\%.


For users to gain privacy during identifier renewal, they must remain offline long enough to mix with other users. Additionally, users must not have a fixed offline time, since this would be susceptible to a timing attack. 
Users must choose an offline time that is not so long to be disruptive, but not so short as to offer little privacy.
The \zipphone population's policy should fix a chosen utility, and employ a cool down time between each user's identifier renewal based on that desired utility. For example, if users' offline-times are 30 seconds, and are aiming to maintain 95\% utility, they will keep every identity for at least $30\textrm{ seconds} \div (1-0.95) = $10~minutes. 

Looking further at the \phl data, we discovered that phone calls were intermittent and could be kept online without sacrificing privacy, though users would miss 4 calls out of 24 per month on average at 95\% utility. However, as we show in Figure~\ref{fig:dark_times}~(right), users cannot remain online during all of their reported interactive time without sacrificing additional privacy. Finally, applications that require a persistent connection to the Internet could not be used without disruption.




\subsection{Battery Overhead}

\zipphone triggers periodic disassociation/association from the mobile carrier, which together incur additional battery draw on the mobile device. Thus, in this section, we quantify  the battery overhead incurred by \zipphone on 3G and 4G networks.

\begin{figure}[t]
\centering
  \includegraphics[width=0.33\textwidth]{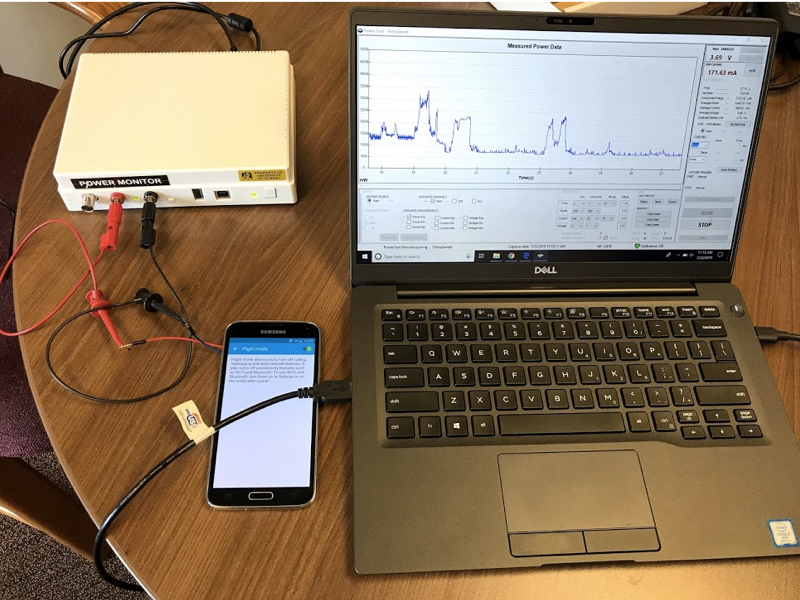}
	\caption{Experimental setup for power measurements on 3G and 4G networks.}\label{fig:powermeter}
\end{figure}
\begin{figure}[t]
	\centering
	\includegraphics[width=0.25\textwidth]{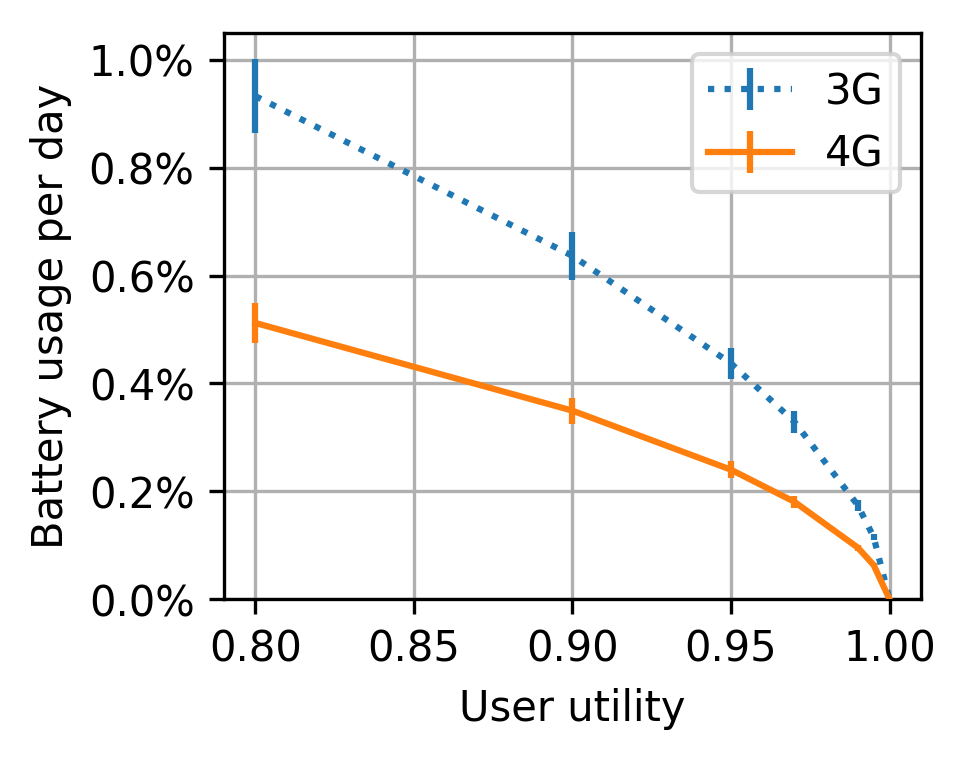}
	\caption{Battery usage does not exceed 1\% per day, regardless of desired privacy or network type.}\label{fig:battuse}
\end{figure}

\textbf{Experimental setup.}
In order to evaluate the power consumption of mobile network association/disassociation, we used a Samsung Galaxy S5 Duos phone with a bypassed battery and a Google Fi SIM card, and a Monsoon Power Meter. We connected the phone to the main channel of the power meter, as illustrated in Figure~\ref{fig:powermeter}, which allowed us to both power up the phone and measure its energy consumption. In order to measure the power draw at 3G and 4G networks, we forced the phone to the respective technology and sampled the power draw at a granularty of 200$\mu$s. We used the phone's Settings screen to toggle between Airplane Mode OFF and Airplane Mode ON every 10 seconds for 4G and every 20 seconds for 3G. We disabled all background services on the phone. This ensured that we are only measuring the power draw from association/disassociation, plus a baseline of about 700mW used by the display for the Airplane Settings page.  For each of 3G and 4G we completed 10 full association/disassociation cycles. The average experienced time and power to connect inform our simulaiton results. 

\begin{figure}[t]
\centering
  \includegraphics[width=0.24\textwidth]{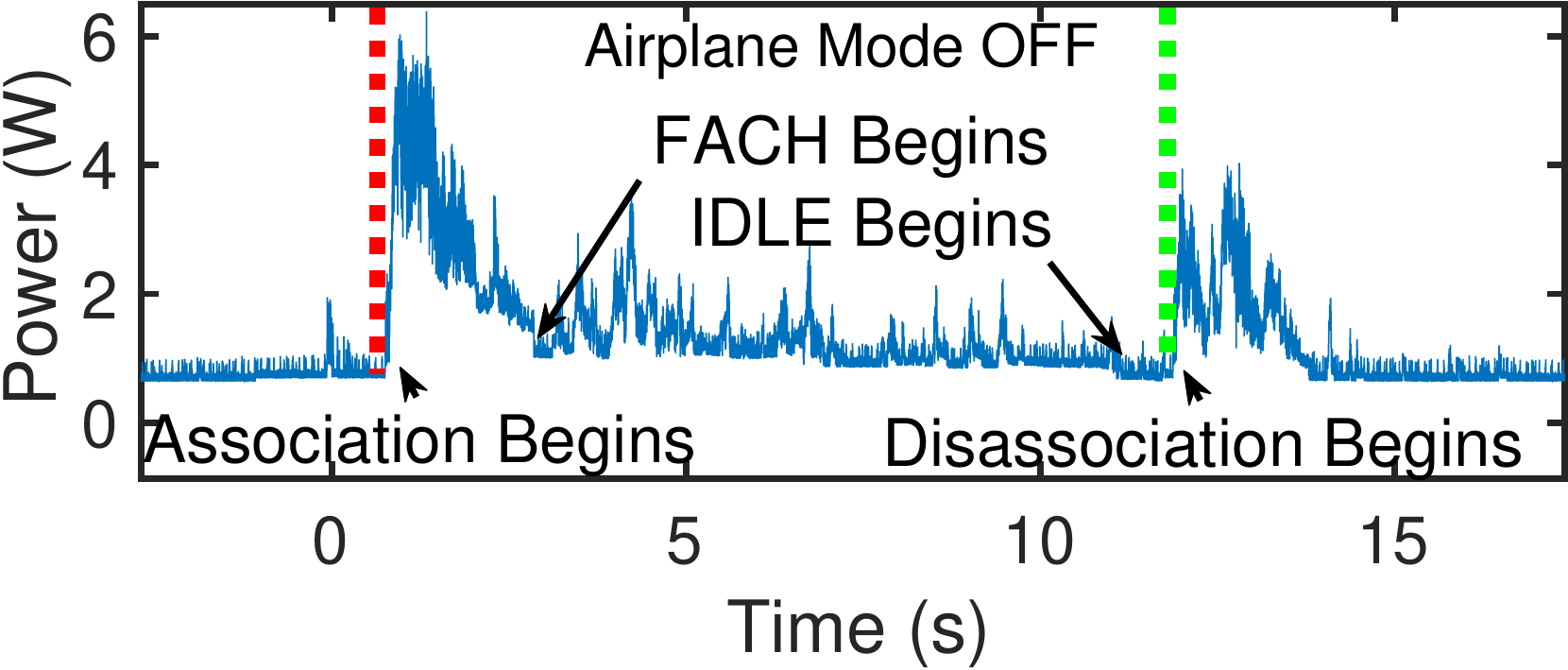}\label{fig:3Gpow}
  \includegraphics[width=0.24\textwidth]{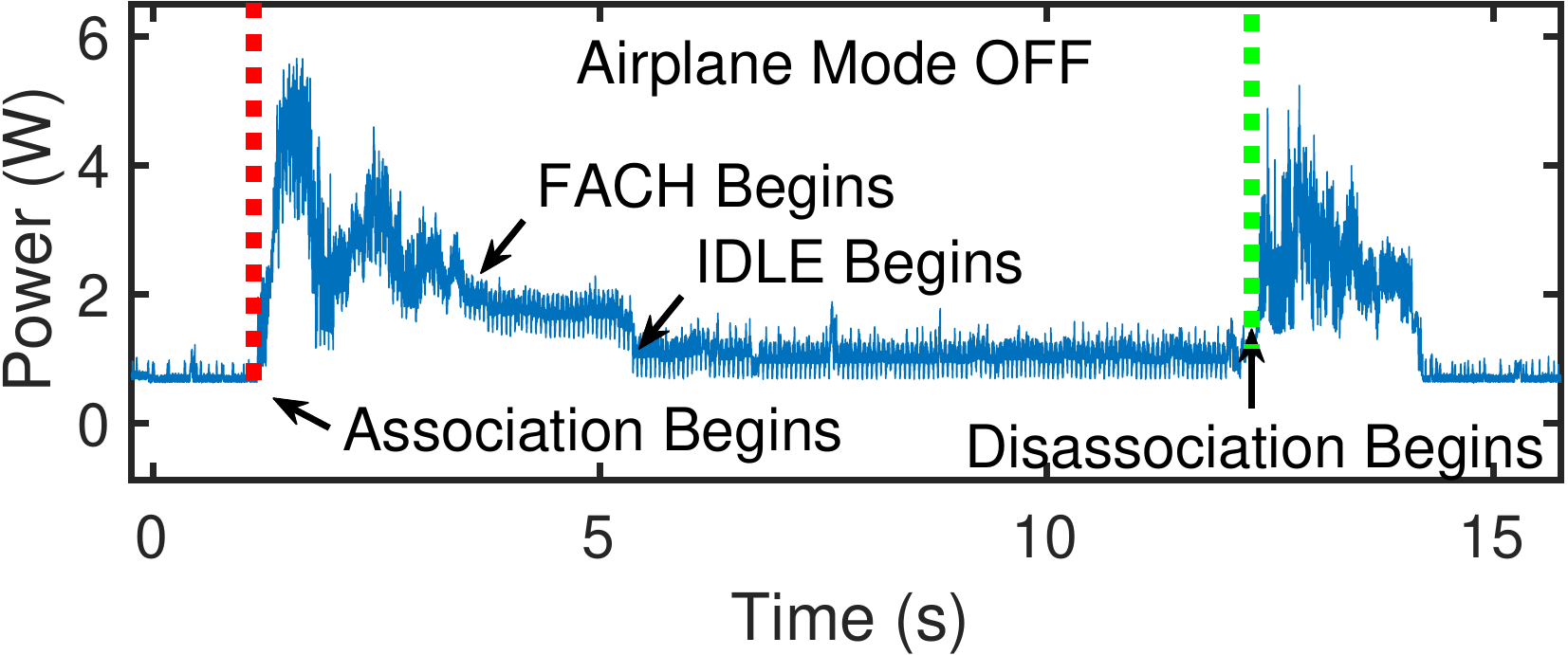}\label{fig:4Gpow}
  \vspace{-0.8cm}
	\caption{Power trace for 3G (left) and 4G (right) association and disassociation.}\label{fig:power}
\end{figure}

Figure~\ref{fig:power} presents a zoomed version of a single associate/disassociate cycle for 3G (left) and 4G (right)\footnote{Note that the timescale (i.e. the $x$-axis range) for 3G is longer than that for 4G, because on 3G the phone takes significantly longer to transition to IDLE mode compared to 4G. }. There are several important points to note on each trace. First, the red vertical line indicates the phone's transition from Airplane Mode ON to OFF state, which immediately triggers a network association. After the association procedure completes, the phone enters FACH (Forward Access CHannel) state in anticipation for the user to begin accessing the Internet. Since this does not happen in our controlled activity, the phone futher transitions into IDLE state. At the instant designated with a green vertical line, we toggle Airplane Mode ON, which immediately triggers a disassociation procedure. 

A \zipphone user would experience two types of overhead: \1~offline time, and \2~power draw. We measure the offline time as the time between the beginning of network association and the beginning of the FACH state. We measure the power overhead as the sum of power to associate and power to disassociate, whereby the power to associate is incurred from the begining of the network association to the beginning of the FACH state, while the power to disassociate is measured from the beginning till the end of the disassociation procedure. 

\begin{figure}[t]\relsize{-1}
\centering
\begin{tabular}{llrr}
\toprule
\multicolumn{2}{l}{}   & {\bf mean}   & {\bf (std dev)} \\ 
\cline{3-4} \noalign{\smallskip}
{\multirow{4}{*}{\bf 3G}} & Power to connect (mW)    & 2,098 & (435)    \\ 
                    & Power to disconnect (mW) & 1,282 & (157)    \\ 
                    & Time to connect (s)      & 5.0   & (0.8)     \\
                    & Time to disconnect (s)   & 4.0   & (1.0)     \\ \hline\noalign{\smallskip}
{\multirow{4}{*}{\bf 4G}} & Power to connect (mW)    & 2,006 & (171)    \\ 
                    & Power to disconnect (mW) & 1,120 & (295)   \\ 
                    & Time to connect (s)      & 2.6   & (0.2)     \\
                    & Time to disconnect (s)   & 3.0   & (1.2)     \\
\bottomrule
\end{tabular}\smallskip
\caption{Time and power overhead incurred by a single association/disassociation procedure on 3G and 4G in our experiments.}\label{tab:powerstats}
\end{figure}

Figure~\ref{tab:powerstats} presents the average incurred overhead for our measurement campaign. We see that the offline time incurred by 3G is nearly double that of 4G. The power consumption, on another hand, is comparable across the two technologies. We use these results to quantify the battery usage per day for users in our datasets. To this end, we convert the measured power consumption for a single connect/disconnect from mW to mWh using the values in Figure~\ref{tab:powerstats}. We assume a 3.85V battery with a capacity of 2800mAh, which is typical. On the $x$-axis we control the desired user utility from 0.8 to 1, which effectively controls the amount of network disconnect/connect cycles a user will incur for the duration of a day. We multiply that number by the energy consumption (in mWh) and then divide by the battery's capacity to determine what fraction of the battery is consumed due to \zipphone. Figure~\ref{fig:battuse} presents our results, which indicate that the battery usage is at most 1\% per day regardless of technology (3G or 4G) or desired privacy. 



%% file: zipphone.tex
\section{Provisioning, Purchase, and Communication}\label{sec:lte}

In this section, we detail how recently standardized  mechanisms for user provisioning and authentication allow for  
\zipphone to be deployed by an untrusted MNO or MVNO.

\subsection{Mobile user provisioning and authentication}
Traditionally, SIM cards  installed in devices provide the basis of  MNO  provisioning. Each SIM has a unique International Mobile Subscriber Identity (IMSI). At the Home Location Registry (HLR) an entry is created connecting the IMSI with an Mobile Station International Subscriber Directory Number (MSISDN; i.e., a phone number).  At the Authentication Center (AuC) the IMSI is paired with a K$_\textrm{i}$ value and used for user authentication. 

This approach is plagued with high overhead, wasted allocations, and manual processes.
 To address these limitations, the   eSIMs standard~\cite{esim_gsmawp} has been developed,
 which allows programmatic and on-the-fly provisioning of a user's identity on a network. With eSIMs, mobile users can maintain multiple simultaneous mobile network identities and use heterogeneous services from one or multiple MNOs. 
Three out of the four major carriers in the US currently support eSIM, with one major carrier supporting eSIM in 42 other countries worldwide~\cite{esim_support}.

eSIMs introduce new components to user management that are useful for \zipphone. Similar to traditional SIMs, the eSIM functional {\em profile}~\cite{simalliance} is jointly maintained in the MNO's HLR and the AuC. The Subscription Manager Data Preparation (SM-DP+), is responsible for provisioning a user's profile onto the eSIM. The SM-DP+ is the first point of communication between an aspiring subscriber and the MNO, from which the subscriber gets their functional profile. There is no upper limit on the amount of \textit{profiles} an eSIM can maintain; this number depends on \1 the size of a single \textit{profile}, \2 the eSIM integrated memory and, \3 the operator's preferences. As an example, T-Mobile currently supports up to 10 concurrent eSIM \textit{Profiles}~\cite{tmobile_profiles}.

Thus, \zipphone can be deployed as an app that anonymously acquires multiple \textit{profiles} from a provider's SM-DP+ and programmatically swaps them in order to improve the subscriber's privacy. Note that this operation leverages the existing MNO mechanisms and, thus does not assume or require any cooperation from the MNO. 
 





%

\subsection{Purchasing Credentials} 
\zipphone requires that users anonymously  purchase  \textit{profiles} without  linking to a consistent financial or network identifier.
{\em Zcash}~\cite{Sasson:2014,Hopwood:2019} offers a basis for such purchases as follows.  The MVNO advertises a Zcash address for purchasing credentials and a price per purchased profile. The user creates a public-private key pair, and issues a transaction that transfers the purchase amount to the MVNO, and includes its own public key as part of the transaction (which could be encrypted with the public key of MVNO's address for greater security). The MVNO responds with a data only transaction that includes the set of credentials, encrypted with the provided public key. This exchange must occur separately for each credential purchased, each with a different public key and each from a different Zcash source address, so that a series of purchases cannot be linked as belonging to the same user. These requirements are easy to provide with Zcash and can be performed programmatically. There is an effort to deploy Zcash's zero-knowledge mechanisms on Ethereum as well.  

The user's transactions should not be overtly issued from the same IP address, as that would allow linking of the purchases. Protocols such as Dandelion++~\cite{Fanti:2018} allow transactions to be issued to Zcash with network anonymity. 

A \zipphone users would carry a series of pre-purchased profiles, and use them in a random order.  MVNO should issue signaling to the MNO to cancel the IMSI  a period of time after they are first used (e.g., 15--30 minutes) .

\subsection{Communication without Leaking Identity or Location}
\label{sec:comm}

\zipphone uses should ignore the MSISDN (phone numbers) provided by a profile. If a \zipphone user initiated or received overt LTE or unencrypted VOIP calls, they risk being identified via a profile of call records held by the carrier. 
expect to retain location privacy. 
(Note that E911 service, which is tied to a handset and not a user or SIM, would be still available if needed.)

Some protection would be gained from using an encrypted VOIP service, since it would not reveal to the carrier the identity of the user's contact, whom she calls, or from whom she receives calls. However, if the IP address of the VOIP service is unique, then connecting to it would help the MNO link a collection of profiles together. An anonymous VOIP service,  such as Torfone can be used; note that anonymous VOIP has a performance penalty~\cite{Liberatore:2011}.

In general, an anonymous communication systems, i.e., Tor,  must be used for all \zipphone communication (voice or data). However, there is  one change required. Tor chooses a consistent, single {\em guard} relay to start all three-relay circuits through the Tor network. If \zipphone users send all traffic to a single guard relays, it would be a consistent identifier despite changing IMSIs. Instead of a guard at the start of the circuit, \zipphone users should use a consistent relay as the exit. This switching of roles allows \zipphone users to receive all protections against the Predecessor Attack~\cite{wright:2004} that Tor normally provides via guard nodes at the entry.

%% file: discussion.tex
\section{Discussion}

\subsubsection{Limitations}\label{sec:limitations}
Our technique has limitations. We require devices that accept software SIMs; these are not common in the marketplace now, though easy to provide. Another limitation is that users would never be able to quantify their privacy gains as there is no way to determine the number of other ZipPhone users. And we do not address other privacy risks, which  include physical attacks (e.g., radio frequency fingerprinting~\cite{deng2017radio}),  software vulnerabilities, use of location-based services, advertising fingerprints, browser cookies, and malware.

Our evaluations are limited as well. For example, we do not explicitly consider users ready to mix when they are stationary; if they do, attackers could also consider these additional mixes when linking. 
A more advanced attacker's classifier might account for yet additional features (e.g., time of day or favourite locations~\cite{Zang:2011}) to increase accuracy. 
Conversely, users could develop more efficacious methods to prevent linking.
And our results are tied to our datasets, which are relatively small and limited to populations from universities. Obtaining a usable dataset is difficult. MNOs are generally unwilling to anonymize and share such data, and collecting data requires a fairly involved longitudinal study. 

Despite these limitations, this paper introduces an effective method for a service provider to put location privacy in the hands of users, and provides a detailed look at the efficacy of such a service.








\subsubsection{Ethical implications}
Mobile devices are an essential part of most people's daily routine. Accordingly, there is a tension between the right to location privacy and the need to investigate crimes and threats to public safety. The techniques we introduce and evaluate are effective to protecting privacy, but unfortunately would thwart a common method of investigation as well. Any deployment of ZipPhone would have to take into account this difficult, zero-sum game ethical dilemma.  



%% file: conclusion.tex
\section{Conclusion}

\zipphone is a novel protocol that provides users increased location privacy while using the existing centralized cellular infrastructure.
We evaluated a deanonymization attack that uses a combination of location profiling and trajectory linking, and showed that it is effective in identifying long-term pseudonyms.
Using two separate datasets of call detail records, we then demonstrated that a \zipphone user can defend against such attacks by renewing her identifier regularly. We also evaluated the utility cost in terms of time offline and battery life, and showed it to be minimal.

Our work demonstrates that, fundamentally, users do not need to trust wireless service providers with their location information. Users who do not use any anonymization scheme are always identifiable. In our trace-driven evaluations, a non-\zipphone user who is habitual and conventional (predictable and mixing) who renews her pseudonym monthly is identifiable 69\% of the time, and one who uses \zipphone is identifiable 24\% of the time if she sacrifices 5\% of her utility and 1\% of battery life, towards a lower bound of 19\% if she sacrifices more. In other words, users can significantly reduce their identifiability by up to 45\% by renewing their pseudonym after an offline period of 30 seconds every ten minutes.



%% file: infocom.bbl
\begin{thebibliography}{10}
\providecommand{\url}[1]{#1}
\csname url@samestyle\endcsname
\providecommand{\newblock}{\relax}
\providecommand{\bibinfo}[2]{#2}
\providecommand{\BIBentrySTDinterwordspacing}{\spaceskip=0pt\relax}
\providecommand{\BIBentryALTinterwordstretchfactor}{4}
\providecommand{\BIBentryALTinterwordspacing}{\spaceskip=\fontdimen2\font plus
\BIBentryALTinterwordstretchfactor\fontdimen3\font minus
  \fontdimen4\font\relax}
\providecommand{\BIBforeignlanguage}[2]{{%
\expandafter\ifx\csname l@#1\endcsname\relax
\typeout{** WARNING: IEEEtran.bst: No hyphenation pattern has been}%
\typeout{** loaded for the language `#1'. Using the pattern for}%
\typeout{** the default language instead.}%
\else
\language=\csname l@#1\endcsname
\fi
#2}}
\providecommand{\BIBdecl}{\relax}
\BIBdecl

\bibitem{Stryjak:2019}
J.~Stryjak and M.~Sivakumaran, ``{The Mobile Economy 2019},'' GMSA
  Intelligence, Tech. Rep., February 25 2019.

\bibitem{Zang:2011}
H.~Zang and J.~Bolot, ``{Anonymization of Location Data Does Not Work: A
  Large-scale Measurement Study},'' in \emph{Proc.\ ACM MobiCom}, 2011, pp.
  145--156.

\bibitem{Mulder:2008}
Y.~D. Mulder, G.~Danezis, L.~Batina, and B.~Preneel, ``{Identification via
  Location-profiling in GSM Networks},'' in \emph{Proc.\ ACM Wrkshp on Privacy
  in the Electronic Society}, 2008, pp. 23--32.

\bibitem{Dingledine:2004}
\BIBentryALTinterwordspacing
R.~Dingledine, N.~Mathewson, and P.~Syverson, ``Tor: {The} second-generation
  onion router,'' in \emph{USENIX Security}, 2004. [Online]. Available:
  \url{https://www.usenix.org/conference/13th-usenix-security-symposium/tor-second-generation-onion-router}
\BIBentrySTDinterwordspacing

\bibitem{niu2015enhancing}
B.~Niu, Q.~Li, X.~Zhu, G.~Cao, and H.~Li, ``Enhancing privacy through caching
  in location-based services,'' in \emph{2015 IEEE conference on computer
  communications (INFOCOM)}.\hskip 1em plus 0.5em minus 0.4em\relax IEEE, 2015,
  pp. 1017--1025.

\bibitem{eff-att}
{Case No. 19-cv-4063}, ``{Scott, Jewel, And Pontis, et al. v. AT\&T Inc.; AT\&T
  Services, Inc.; AT\&T Mobility, LLC; Technocom Corp.; and Zumigo, Inc.}''
  \url{https://www.courthousenews.com/wp-content/uploads/2019/07/ATTlocationservices-COMPLAINT.pdf},
  July 2019.

\bibitem{Beresford:2003}
A.~Beresford and F.~Stajano, ``Location privacy in pervasive computing,''
  \emph{IEEE Pervasive Computing}, vol.~2, no.~1, pp. 46--55, 2003.

\bibitem{Eagle:2006}
N.~Eagle and A.~Pentland, ``{Reality Mining: Sensing Complex Social Systems},''
  \emph{Personal and Ubiquitous Computing}, vol.~10, no.~4, pp. 255--268, 2006.

\bibitem{nandugudi2013phonelab}
A.~Nandugudi, A.~Maiti, T.~Ki, F.~Bulut, M.~Demirbas, T.~Kosar, C.~Qiao, S.~Y.
  Ko, and G.~Challen, ``Phonelab: A large programmable smartphone testbed,'' in
  \emph{Proceedings of First International Workshop on Sensing and Big Data
  Mining}.\hskip 1em plus 0.5em minus 0.4em\relax ACM, 2013, pp. 1--6.

\bibitem{de2013unique}
Y.-A. De~Montjoye, C.~A. Hidalgo, M.~Verleysen, and V.~D. Blondel, ``Unique in
  the crowd: The privacy bounds of human mobility,'' \emph{Scientific reports},
  vol.~3, p. 1376, 2013.

\bibitem{yin2015re}
L.~Yin, Q.~Wang, S.-L. Shaw, Z.~Fang, J.~Hu, Y.~Tao, and W.~Wang,
  ``Re-identification risk versus data utility for aggregated mobility research
  using mobile phone location data,'' \emph{PloS one}, vol.~10, no.~10, p.
  e0140589, 2015.

\bibitem{tu2018new}
Z.~Tu, F.~Xu, Y.~Li, P.~Zhang, and D.~Jin, ``A new privacy breach: User
  trajectory recovery from aggregated mobility data,'' \emph{IEEE/ACM
  Transactions on Networking}, vol.~26, no.~3, pp. 1446--1459, 2018.

\bibitem{gramaglia2017preserving}
M.~Gramaglia, M.~Fiore, A.~Tarable, and A.~Banchs, ``Preserving mobile
  subscriber privacy in open datasets of spatiotemporal trajectories,'' in
  \emph{IEEE INFOCOM 2017-IEEE Conference on Computer Communications}.\hskip
  1em plus 0.5em minus 0.4em\relax IEEE, 2017, pp. 1--9.

\bibitem{dwork2011differential}
C.~Dwork, ``Differential privacy,'' \emph{Encyclopedia of Cryptography and
  Security}, pp. 338--340, 2011.

\bibitem{mir2013dp}
D.~J. Mir, S.~Isaacman, R.~C{\'a}ceres, M.~Martonosi, and R.~N. Wright,
  ``Dp-where: Differentially private modeling of human mobility,'' in
  \emph{2013 IEEE international conference on big data}.\hskip 1em plus 0.5em
  minus 0.4em\relax IEEE, 2013, pp. 580--588.

\bibitem{Isaacman:2011}
S.~Isaacman, R.~Becker, R.~C\'{a}ceres, S.~Kobourov, M.~Martonosi, J.~Rowland,
  and A.~Varshavsky, ``{Identifying Important Places in People's Lives from
  Cellular Network Data},'' in \emph{Proc.\ Intl.\ Conf.\ on Pervasive
  Computing}, 2011, pp. 133--151.

\bibitem{Beresford:2004}
A.~Beresford and F.~Stajano, ``Mix zones: user privacy in location-aware
  services,'' in \emph{Proc.\ Pervasive Computing and Communications Wrkshps},
  2004, pp. 127--131.

\bibitem{Krumm:2007}
J.~Krumm, ``{Inference Attacks on Location Tracks},'' in \emph{Proc.\ Intl.\
  Conf.\ on Pervasive Computing}, May 2007, pp. 127--143.

\bibitem{Golle:2009}
P.~Golle and K.~Partridge, ``On the anonymity of home/work location pairs,'' in
  \emph{Proc.\ Intl.\ Conf.\ on Pervasive Computing}, 2009, pp. 390--397.

\bibitem{Ma:2010}
C.~Y. Ma, D.~K. Yau, N.~K. Yip, and N.~S. Rao, ``Privacy vulnerability of
  published anonymous mobility traces,'' in \emph{Proc.\ MobiCom}, 2010, pp.
  185--196.

\bibitem{huang2018location}
H.~Huang, G.~Gartner, J.~M. Krisp, M.~Raubal, and N.~Van~de Weghe, ``Location
  based services: ongoing evolution and research agenda,'' \emph{Journal of
  Location Based Services}, vol.~12, no.~2, pp. 63--93, 2018.

\bibitem{wang2018protecting}
J.~Wang, Z.~Cai, Y.~Li, D.~Yang, J.~Li, and H.~Gao, ``Protecting query privacy
  with differentially private k-anonymity in location-based services,''
  \emph{Personal and Ubiquitous Computing}, vol.~22, no.~3, pp. 453--469, 2018.

\bibitem{wang2018privacy}
S.~Wang, Q.~Hu, Y.~Sun, and J.~Huang, ``Privacy preservation in location-based
  services,'' \emph{IEEE Communications Magazine}, vol.~56, no.~3, pp.
  134--140, 2018.

\bibitem{hong2018guti}
B.~Hong, S.~Bae, and Y.~Kim, ``Guti reallocation demystified: Cellular location
  tracking with changing temporary identifier.'' in \emph{NDSS}, 2018.

\bibitem{hussain2019insecure}
S.~R. Hussain, M.~Echeverria, A.~Singla, O.~Chowdhury, and E.~Bertino,
  ``Insecure connection bootstrapping in cellular networks: the root of all
  evil,'' in \emph{Proceedings of the 12th Conference on Security and Privacy
  in Wireless and Mobile Networks}.\hskip 1em plus 0.5em minus 0.4em\relax ACM,
  2019, pp. 1--11.

\bibitem{khan2018defeating}
M.~Khan, P.~Ginzboorg, K.~J{\"a}rvinen, and V.~Niemi, ``Defeating the downgrade
  attack on identity privacy in 5g,'' in \emph{International Conference on
  Research in Security Standardisation}.\hskip 1em plus 0.5em minus 0.4em\relax
  Springer, 2018, pp. 95--119.

\bibitem{Federrath:1996}
H.~Federrath, A.~Jerichow, and A.~Pfitzmann, ``{MIXes in Mobile Communication
  Systems: Location Management with Privacy},'' in \emph{Proc.\ Intl.\ Wrkshp
  on Information Hiding}, 1996, pp. 121--135.

\bibitem{Gorlatova:2011}
M.~Gorlatova, R.~Aiello, and S.~Mangold, ``Managing location privacy in
  cellular networks with femtocell deployments,'' in \emph{Proc.\ WiOpt
  Symposium}, May 2011, pp. 418--422.

\bibitem{Gorlatova:2011a}
------, ``Managing base station location privacy,'' in \emph{Proc.\ MILCOM},
  Nov. 2011, pp. 1201--1206.

\bibitem{Foo:2012}
D.~F. Kune, J.~Koelndorfer, N.~Hopper, and Y.~Kim, ``{Location leaks on the GSM
  Air Interface},'' in \emph{Proc.\ ISOC NDSS}, Feb. 2012.

\bibitem{Reed:1998}
M.~G. Reed, P.~F. Syverson, and D.~M. Goldschlag, ``{Protocols using anonymous
  connections: Mobile applications},'' in \emph{Security Protocols}, ser. LNCS,
  1998, vol. 1361, pp. 13--23.

\bibitem{Federrath:1995}
H.~Federrath, A.~Jerichow, D.~Kesdogan, and A.~Pfitzmann, ``{Security in Public
  Mobile Communication Networks},'' in \emph{Proc.\ IFIP/TC6 Personal Wireless
  Communications}, April 1995, pp. 105--116.

\bibitem{Fatemi:2010}
M.~Fatemi, S.~Salimi, and A.~Salahi, ``Anonymous roaming in universal mobile
  telecommunication system mobile networks,'' \emph{IET Information Security
  Journal}, vol.~4, no.~2, pp. 93--103, 2010.

\bibitem{Park:2001}
J.~Park, J.~Go, and K.~Kim, ``Wireless authentication protocol preserving user
  anonymity,'' in \emph{Proc.\ SCIS}, 2001, pp. 159--164.

\bibitem{Jiang:2006}
Y.~Jiang, C.~Lin, X.~Shen, and M.~Shi, ``{Mutual Authentication and Key
  Exchange Protocols for Roaming Services in Wireless Mobile Networks},''
  \emph{IEEE Trans.\ on Wireless Communications}, vol.~5, no.~9, pp.
  2569--2577, 2006.

\bibitem{Yang:2005}
G.~Yang, D.~Wong, and X.~Deng, ``Efficient anonymous roaming and its security
  analysis,'' in \emph{Applied Cryptography and Network Security}, ser. LNCS,
  2005, vol. 3531, pp. 334--349.

\bibitem{Zhu:2004}
J.~Zhu and J.~Ma, ``A new authentication scheme with anonymity for wireless
  environments,'' \emph{IEEE Trans.\ on Consumer Electronics}, vol.~50, no.~1,
  pp. 231--235, 2004.

\bibitem{Kesdogan:1996}
D.~Kesdogan, H.~Federrath, A.~Jerichow, and A.~Pfitzmann, ``{Location
  Management Strategies Increasing Privacy in Mobile Communication},'' in
  \emph{{Information Systems Security}}, 1996, pp. 39--48.

\bibitem{Freudiger:2009}
J.~Freudiger, R.~Shokri, and J.-P. Hubaux, ``{On the Optimal Placement of Mix
  Zones},'' in \emph{Proc.\ PETS}, Aug. 2009, pp. 216--234.

\bibitem{Bindschaedler:2012}
L.~Bindschaedler, M.~Jadliwala, I.~Bilogrevic, I.~Aad, J.-P. Hubaux, V.~Niemi,
  and P.~Ginzboorg, ``{Track Me If You Can: On the Effectiveness of
  Context-based Identifier Changes in Deployed Mobile Networks},'' in
  \emph{Proc.\ ISOC NDSS}, Feb. 2012.

\bibitem{Shokri:2011}
R.~Shokri, G.~Theodorakopoulos, G.~Danezis, J.-P. Hubaux, and J.-Y. Boudec,
  ``{Quantifying Location Privacy: The Case of Sporadic Location Exposure},''
  in \emph{Proc.\ PETS}, Aug. 2011, pp. 57--76.

\bibitem{Kido:2005}
H.~Kido, Y.~Yanagisawa, and T.~Satoh, ``An anonymous communication technique
  using dummies for location-based services,'' in \emph{Proc.\ Intl.\ Conf.\ on
  Pervasive Services}, 2005, pp. 88--97.

\bibitem{chan2015anoncall}
E.~Chan-Tin, ``Anoncall: Making anonymous cellular phone calls,'' in \emph{2015
  10th International Conference on Availability, Reliability and
  Security}.\hskip 1em plus 0.5em minus 0.4em\relax IEEE, 2015, pp. 626--631.

\bibitem{emara2015caps}
K.~Emara, W.~Woerndl, and J.~Schlichter, ``Caps: Context-aware privacy scheme
  for vanet safety applications,'' in \emph{Proceedings of the 8th ACM
  conference on security \& privacy in wireless and mobile networks}.\hskip 1em
  plus 0.5em minus 0.4em\relax ACM, 2015, p.~21.

\bibitem{Dabrowski:2014}
A.~Dabrowski, N.~Pianta, T.~Klepp, M.~Mulazzani, and E.~Weippl, ``{IMSI-Catch
  Me If You Can: IMSI-Catcher-Catchers},'' in \emph{Proc.\ ACM ACSAC}, 2014.

\bibitem{Corner:2017}
M.~D. Corner, B.~N. Levine, O.~Ismail, and A.~Upreti, ``{Advertising-based
  Measurement: A Platform of 7 Billion Mobile Devices},'' in \emph{ACM
  International Conference on Mobile Computing and Networking (MobiCom)},
  {October} 2017.

\bibitem{silentcircle}
``{Silent Circle} blackphone,'' \url{http://silentcircle.com}.

\bibitem{torfone}
``Torfone,'' \url{http://torfone.org}.

\bibitem{esim_gsmawp}
``{eSIM Whitepaper: The what and how of Remote SIM Provisioning},''
  \url{https://www.gsma.com/esim/wp-content/uploads/2018/12/esim-whitepaper.pdf},
  March 2018.

\bibitem{esim_support}
``{Find wireless carriers that offer eSIM service},''
  \url{https://support.apple.com/en-us/HT209096}, July 2019.

\bibitem{simalliance}
``{eUICC Technical Releases},''
  \url{https://simalliance.org/euicc/euicc-technical-releases/}.

\bibitem{tmobile_profiles}
``{eSIM settings: Apple iPhone on iOS 12},''
  \url{https://support.t-mobile.com/docs/DOC-39253}.

\bibitem{Sasson:2014}
E.~B. {Sasson}, A.~{Chiesa}, C.~{Garman}, M.~{Green}, I.~{Miers}, E.~{Tromer},
  and M.~{Virza}, ``Zerocash: Decentralized anonymous payments from bitcoin,''
  in \emph{2014 IEEE Symposium on Security and Privacy}, May 2014, pp.
  459--474.

\bibitem{Hopwood:2019}
D.~Hopwood, S.~Bowe, T.~Hornby, and N.~Wilcox, ``Zcash protocol specification
  version,'' April 18 2019.

\bibitem{Fanti:2018}
G.~Fanti, S.~B. Venkatakrishnan, S.~Bakshi, B.~Denby, S.~Bhargava, A.~Miller,
  and P.~Viswanath, ``Dandelion++: Lightweight cryptocurrency networking with
  formal anonymity guarantees,'' \emph{Proc. ACM Meas. Anal. Comput. Syst.},
  vol.~2, no.~2, pp. 29:1--29:35, Jun. 2018.

\bibitem{Liberatore:2011}
M.~Liberatore, B.~Gurung, B.~N. Levine, and M.~Wright, ``{Empirical Tests of
  Anonymous Voice Over IP},'' \emph{Elsevier Journal of Network and Computer
  Applications}, vol.~34, no.~1, pp. 341--350, January 2011.

\bibitem{wright:2004}
M.~Wright, M.~Adler, B.~N. Levine, and C.~Shields, ``{The Predecessor Attack:
  An Analysis of a Threat to Anonymous Communications Systems},'' \emph{ACM
  Transactions on Information and System Security (TISSEC)}, vol.~4, no.~7, pp.
  489--522, November 2004.

\bibitem{deng2017radio}
S.~Deng, Z.~Huang, X.~Wang, and G.~Huang, ``Radio frequency fingerprint
  extraction based on multidimension permutation entropy,'' \emph{International
  Journal of Antennas and Propagation}, vol. 2017, 2017.

\end{thebibliography}
